\documentclass[prb,twocolumn,amsmath,amssymb,showpacs]{revtex4-1} 
\usepackage{graphics}
\usepackage{dcolumn}
\usepackage{bm}
\usepackage{color}
\usepackage{epstopdf}
\usepackage{epsfig}

\begin{document}

\title{Field-induced inter-planar correlations in the high-temperature superconductor La$_{1.88}$Sr$_{0.12}$CuO$_4$}

\author{A. T. R\o mer,$^{1}$ P. Jensen,$^1$ H. Jacobsen,$^{1}$ L. Udby,$^1$ B. M. Andersen,$^1$ M. Bertelsen,$^1$ S. L. Holm,$^1$  N. B. Christensen,$^{2}$ 
R. Toft-Petersen,$^3$ M. Skoulatos,$^4$ M. Laver,$^4$ A. Schneidewind,$^5$ P. Link,$^6$ M. Oda,$^{7}$ M. Ido,$^{7}$ N. Momono,$^8$ and K. Lefmann$^{1}$}
\affiliation{$^1$Nanoscience Center, Niels Bohr Institute, University of Copenhagen, DK-2100 Copenhagen, Denmark\\
$^2$Department of Physics, Technical University of Denmark, DK-2800 Kgs. Lyngby, Denmark\\
$^3$Helmholtz Zentrum Berlin f\"ur Materialien und Energie D-14109 Berlin, Germany\\
$^4$Laboratory of Neutron Scattering,  Paul Scherrer Institute, 5232-Villigen, Switzerland\\
$^5$J\"ulich Centre for Neutron science (JCNS) at MLZ, Forschungszentrum J\"ulich GmbH, Outstation MLZ, Lichtenbergstr. 1, 85747 Garching, Germany\\
$^6$Heinz Maier-Leibnitz-Zentrum (MLZ), TU M\"unchen, Lichtenberstr. 1, 85747 Garching, Germany\\
$^7$Department of Physics, Hokkaido University, Sapporo 060-0810, Japan\\
$^8$Department of Applied Sciences, Muroran Institute of Technology, Muroran 050-8585, Japan}

\begin{abstract}
We present neutron scattering studies of the inter-planar correlations in the high-temperature superconductor La$_{1.88}$Sr$_{0.12}$CuO$_4$ ($T_c=27$ K). The correlations are studied both in a magnetic field applied perpendicular to the CuO$_2$ planes, and in zero field under different cooling conditions. We find that the effect of the magnetic field is to increase the magnetic scattering signal at all values of the out-of-plane wave vector $L$, indicating an overall increase of the magnetic moments. In addition, weak correlations between the copper oxide planes develop in the presence of a magnetic field. This effect is not taken into account in previous reports on
the field effect of magnetic scattering, since usually only $L\approx 0$ is probed. Interestingly, the results of quench-cooling the sample are similar to those obtained by applying a magnetic field. Finally, a small variation of the incommensurate peak position as a function of $L$ provides evidence that the incommensurate signal is twinned with the dominating and sub-dominant twin displaying peaks at even or odd $L$, respectively. 
\end{abstract}


\maketitle

\section{Introduction}
The interplay between magnetic ordering and superconductivity remains a topic of intense investigation in both cuprates and iron-based superconductors.~\cite{recentreviews} 
In the single-layer cuprate superconductor 
La$_{2-x}$Sr$_x$CuO$_4$ (LSCO), incommensurate (IC) magnetic order and fluctuations have been 
observed at a quartet of IC positions around the magnetic ordering vector in the parent compound La$_{2}$CuO$_4$ (LCO), i.e. ${\bf Q}_{\rm IC} = (1\pm\delta_H,~\pm\delta_K~,0)$ in orthorhombic notation.\cite{Birgeneau88,Birgeneau89,Kimura00,LSCOreview1,LSCOreview2} In the doping range
$0.06 \leq x \leq 0.13$ it was shown by Yamada {\em et al.}\cite{Yamada98} that the incommensurability $\delta$ scales linearly with the doping, $\delta \approx x$. 
For doping levels close to $x = 0.125$, the superconducting critical transition temperature is somewhat suppressed,~\cite{Katano00} which is known as the ’1/8 anomaly’, and caused by stripe ordering.\cite{LBCO1,LBCO2,LBCO3}
The static magnetism in LSCO near $x=0.125$, as well as its momentum space characteristics has been previously studied in great detail.~\cite{Suzuki98,Kimura00,Katano00}

Several experiments have shown that application of a magnetic field perpendicular to the CuO$_2$ planes leads to an enhancement of the elastic response from the magnetic IC order at ${\bf Q}_{\rm IC}$ for doping values in a range around the 1/8 anomaly: $0.10 \leq x \leq 0.135$.~\cite{Lake02,Chang08,Kofu09} In LSCO of higher doping no static order is present, but it has been shown that magnetic order can be induced by application of a magnetic field.~\cite{Chang09,Khaykovich}

Earlier, some of us studied the magnetic correlations along the $c$-axis in a crystal with doping value of $x=0.11$.~\cite{Lake05,RealdopingofBellaCryst} We observed that only the magnetic field component perpendicular to the superconducting $(a,b)$ plane gives rise to an enhanced IC scattering, whereas the field component in the plane does not affect the magnetic IC signal. Further, the field-induced intensity was modulated along the $c$-axis, indicating that inter-planar spin correlations develop in the presence of a magnetic field perpendicular to the CuO$_2$ planes. 

The observation of enhanced $c$-axis correlations raises a concern: Parts of, or in principle all of, the field-induced signal observed in measurements using the more common $(a,b)$ plane crystal orientation may be due to the induced correlations, and not to an increase of static magnetism in the superconductor, as commonly believed.
In the present work, we perform a comprehensive study of the field-induced signal of a LSCO crystal of a slightly larger doping level, $x=0.12$. Our results confirm the earlier findings of both field-induced magnetism and field-induced $c$-axis correlations in this system. In particular, we find that much of the observed IC signal in our experiments arises from an actual increase of magnetism in the system. In addition, short range $c$-axis correlations develop. We present estimates for corrections of the values of field-induced signal arising due to $c$-axis correlations. Surprisingly, we also find that fast cooling of the crystal to base temperature induces short range $c$-axis correlations similar to what is found when applying a strong magnetic field. In combination with observations by Lee {\it et al.}\cite{Lee04} on oxygen-doped La$_2$CuO$_4$, these observations suggest that fast cooling and application of a magnetic field have similar effects on the IC order.
\begin{figure*}[t!]
\centering
\includegraphics[clip=true,width=0.85\columnwidth]{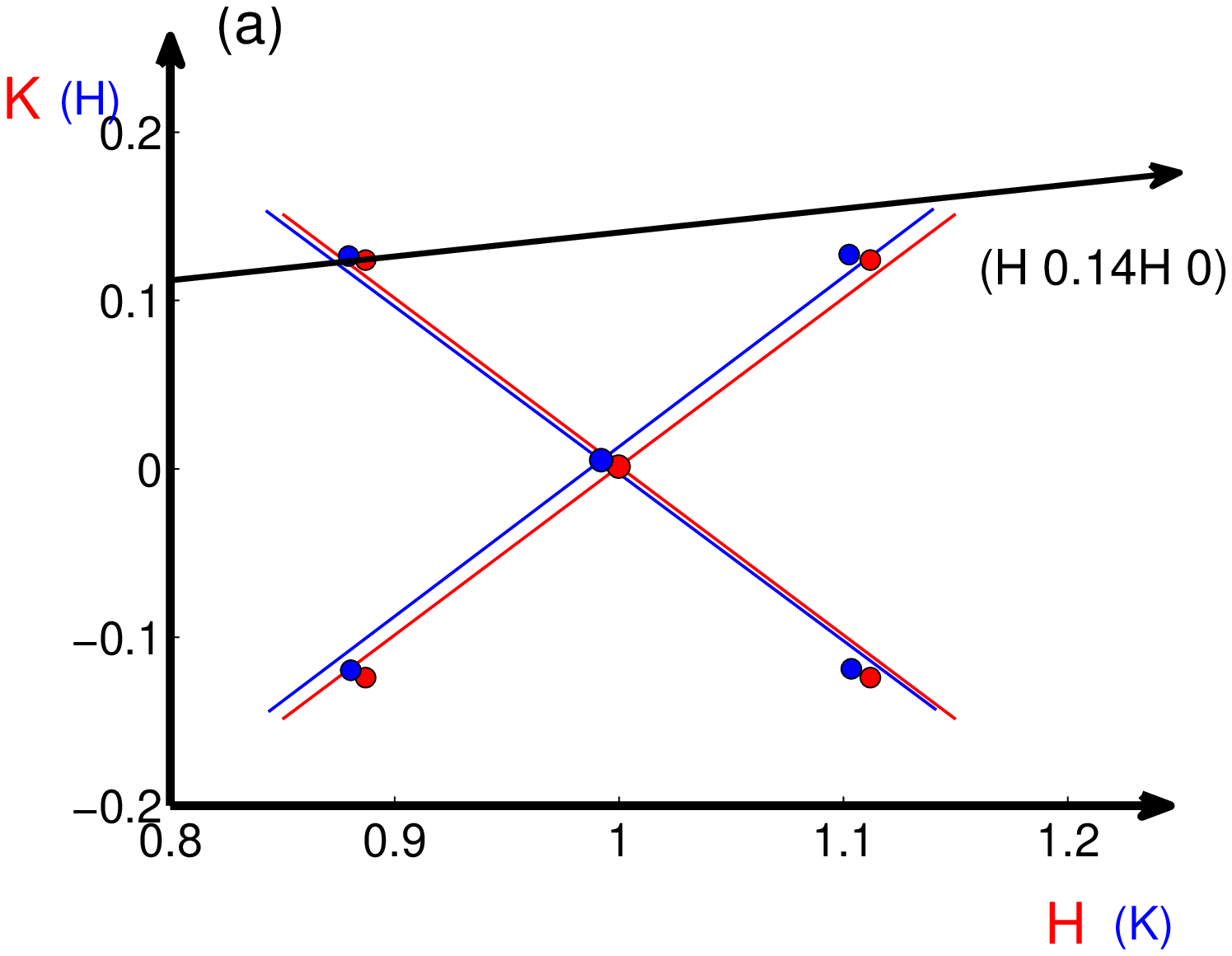}
\includegraphics[clip=true,width=0.85\columnwidth]{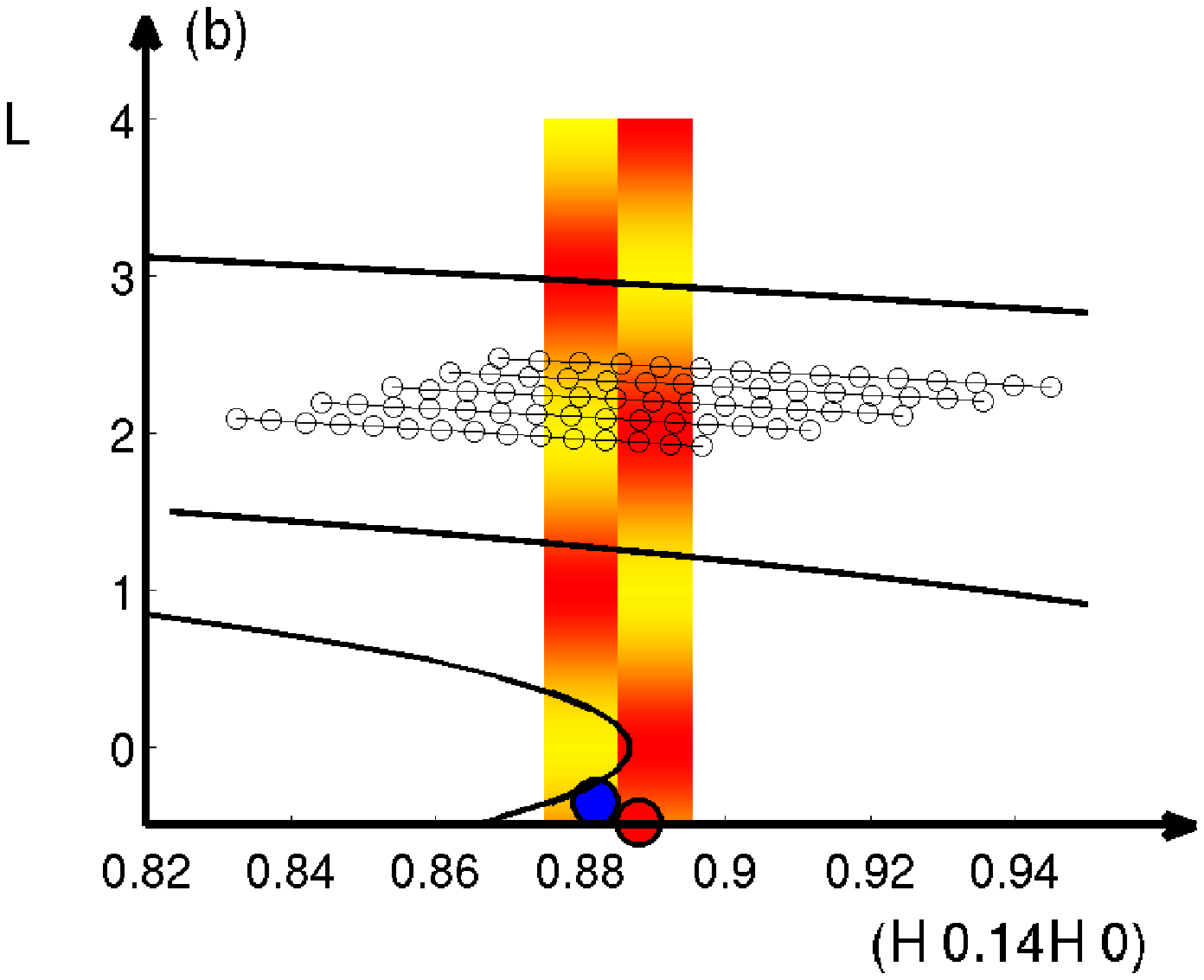}
\caption{(color online) (a) Illustration of the quartet of IC peaks around $(1~0~0)$ in orthorhombic notation. The scattering signal from the $(1~0~0)$ structural second order peak belonging to the dominant domain is shown by the red circle. The crystal shows twinning into two sub-dominant domains of which we show the strongest by the blue circles. The red and blue lines show the Cu-O-Cu axes. Note that the IC peaks are shifted off these high symmetry axes in agreement with earlier reports.~\cite{Kimura00} The black arrow shows the axis $(H ~0.14H~0)$ which is in the scattering plane. (b) Illustration of the scattering plane spanned by $(H ~0.14H~0)$ and $(0~0~L)$. We show examples of a sample rotation scan through a signal rod with a proposed weak $L$ dependence of the signal shown for the two twins of the IC signal, with maximum signal visualized by 
red color and minimum signal shown in yellow. The width of the signals along $(H ~0.14H~0)$ is exaggerated for clarity. Note that one twin (red circle) displays peaks at even $L$ values and the twin shown by the 
blue circle exhibits peaks at odd 
$L$ values. The white points visualize how the 9 analyzer blades 
of imaging mode on RITA-II enable 
measurements at distinct values of $L$ for one sample rotation scan. For clarity, only every second analyzer blade has been shown. The solid lines show the dependence of the reciprocal lattice vectors in a sample rotation scan. From the scan line centered at $L=0$, it is clear that a sample rotation scan is not applicable for small $L$-values. With increasing $L$, the change in $L$ during one sample rotation scan becomes smaller.}
\label{fig:a3scan}
\end{figure*}

\section{Experimental method}
\label{section_exp}
The La$_{1.88}$Sr$_{0.12}$CuO$_4$ ($T_{\rm c} = 27\pm1.5$ K) sample studied in this work consisted of a single crystal grown
by the traveling solvent floating zone method.~\cite{Nakano98} In earlier work on the same crystal,~\cite{Chang08,GilardiThesis} the Sr content $x=0.120 \pm0.005$ was determined from the structural transition temperature separating the high-temperature tetragonal (HTT) from the low-temperature orthorhombic (LTO) phase. A neutron diffraction scan of the structural $(2~0~0)$ reflection shows that the crystal displays twinning into primarily two domains. We have checked that the twin pattern is reproducible under slow cooling conditions. 

High-resolution elastic neutron scattering experiments were carried out on three different cold neutron triple axis spectrometers: RITA-II\cite{RITAinstr} at the SINQ neutron source at PSI, Switzerland, FLEXX\cite{MDLe13}  at the BER2 research reactor at HZB Berlin, Germany, and PANDA\cite{PANDA} at the FRM-II research reactor source in Munich, Germany. Preliminary data were taken at BT-7,\cite{Lynn12} NIST. 
We perform measurements with the sample oriented with the $c$-axis in the scattering plane. To obtain scattering from an IC position we therefore tilt the $(a,b)$ plane $\sim7.8^\circ$ out of the scattering plane as illustrated in Fig.~\ref{fig:a3scan}. Thereby we get access to wave vectors of the form ${\bf Q}=(H~0.14H~L)$ in orthorhombic notation, where $a=5.312$ \AA, $b=5.356$ \AA, $c=13.229$ \AA. We find an IC signal at the position ${\bf Q_{\rm IC}}=(0.887(2),0.124(1),L)$.

In triple axis spectrometers, the resolution ellipsoid is elongated out of the scattering plane. This means that in common experiments, where the $(a,b)$ plane is in the scattering plane, the intensity is enhanced by resolution integration along the $c$-axis, along which the IC signal is broad. In the present crystal alignment with the $c$-axis in the scattering plane we do not gain intensity by these resolution effects and optimization of the experimental setup is required. Sample rotation scans are optimal in this situation since this limits distinct background contributions from e.g. powder lines arising from 2nd and 3rd 
order neutrons reflected by the monochromator. All experiments were therefore carried out by sample rotation scans with the exception of a few scans close to $L=0$ in the zero field experiment at PANDA, see the white circles and solid black lines of Fig.~\ref{fig:a3scan} (b) for an illustration of sample rotation scans.

\begin{figure*}
\centering
\includegraphics[clip=true,width=0.32\linewidth]{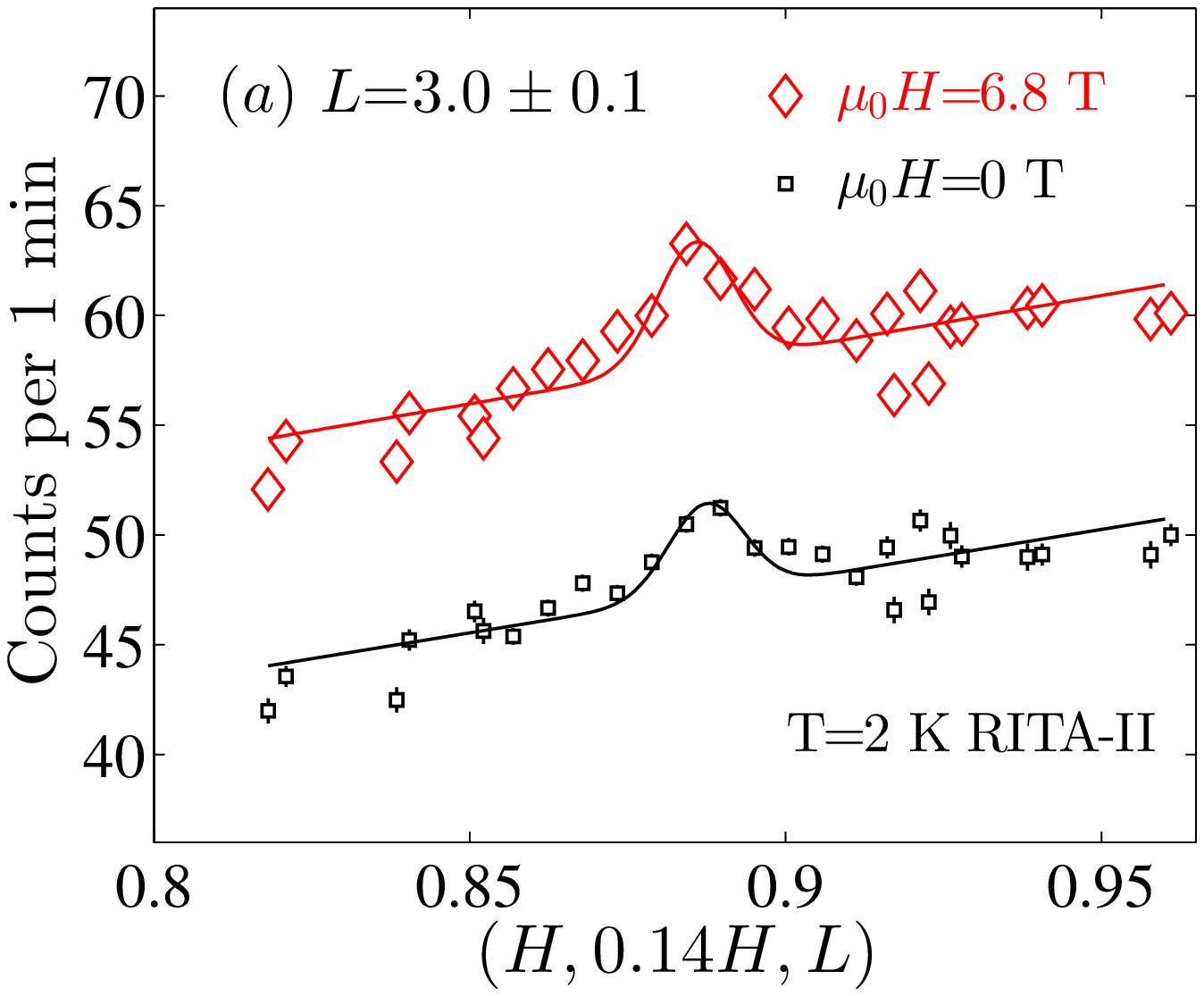}
\includegraphics[clip=true,width=0.32\linewidth]{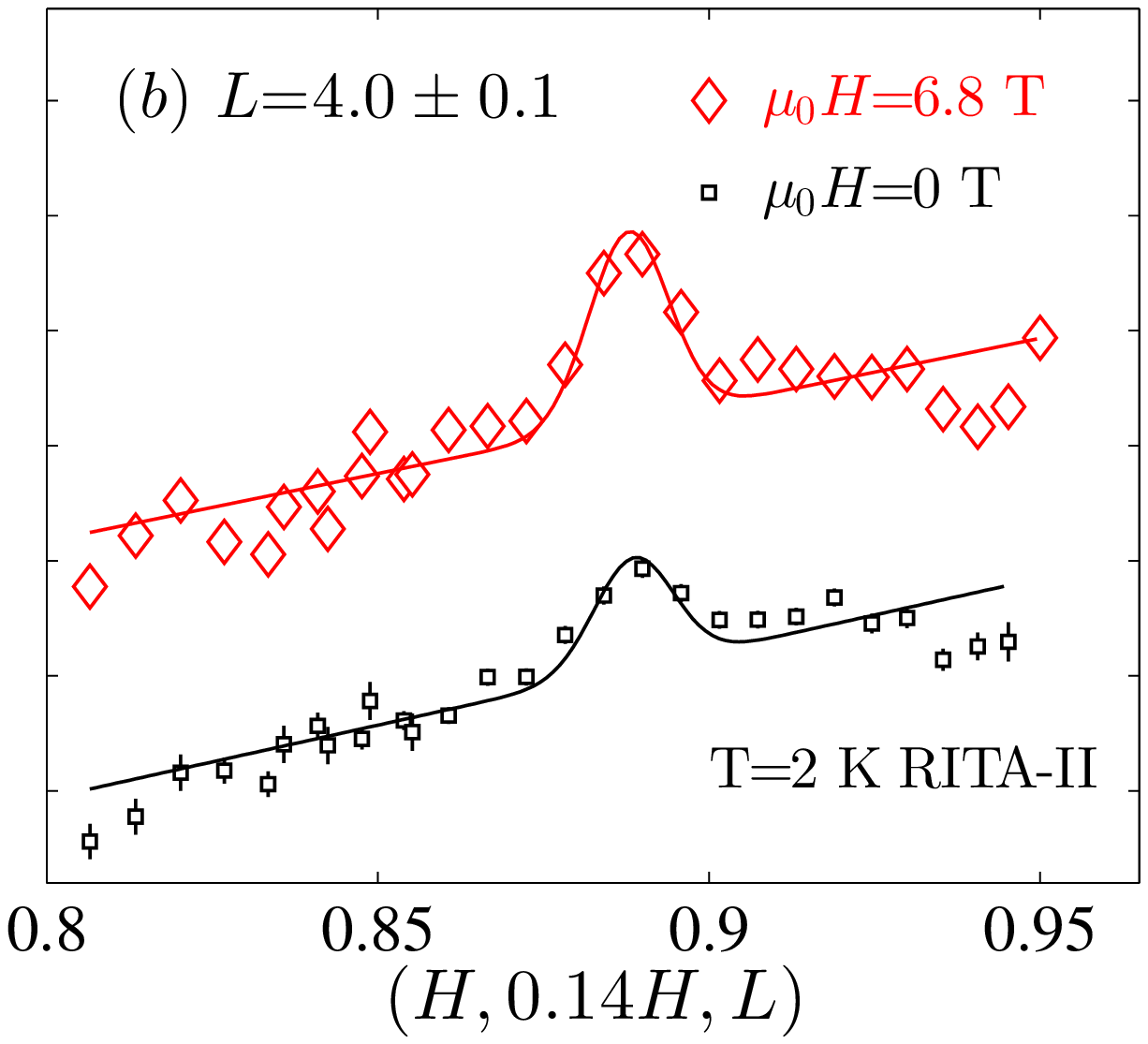}
\includegraphics[clip=true,width=0.32\linewidth]{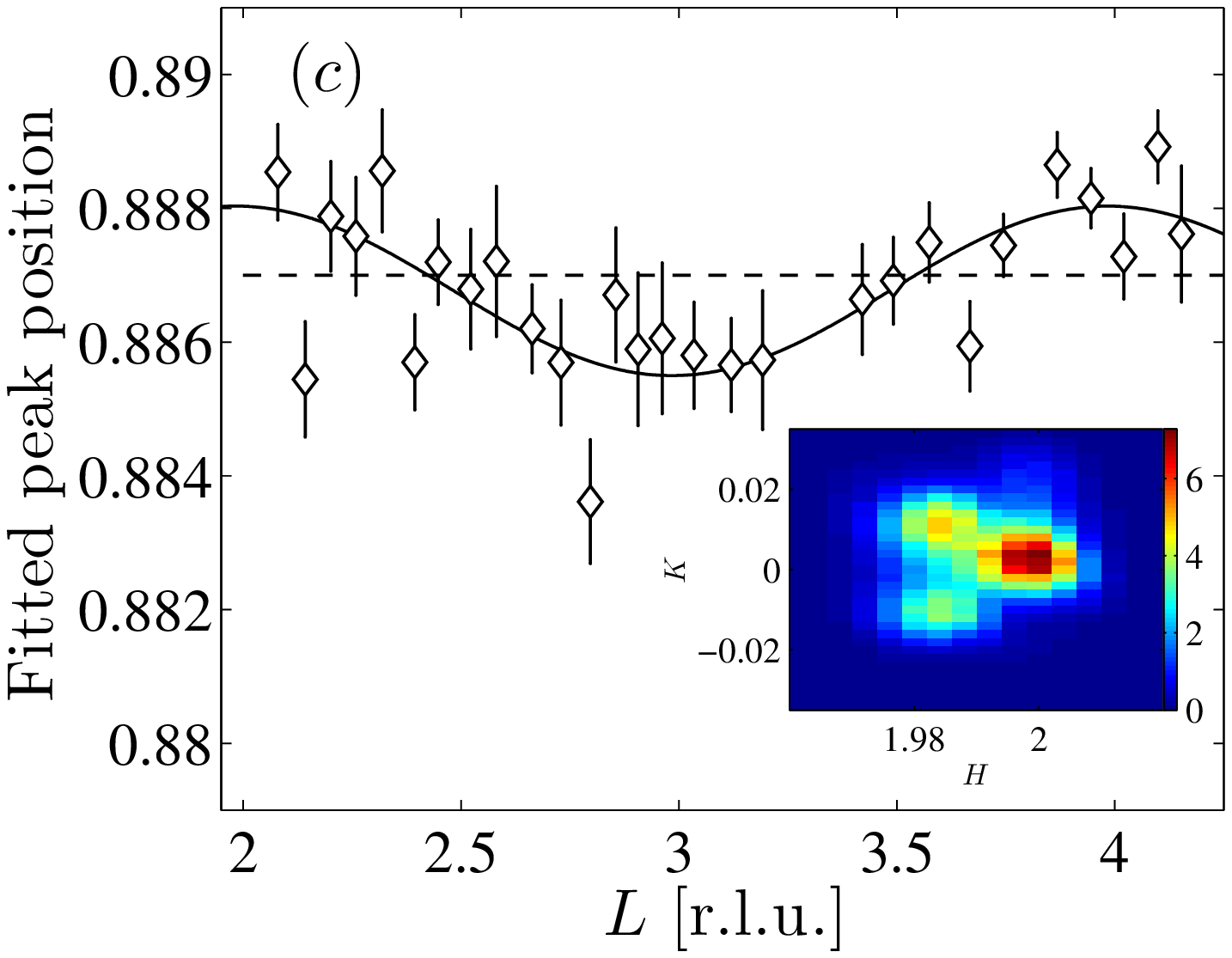}
\caption{(color online) (a-b) Raw data for sample rotation scans through the IC position $(H~0.14H~L)$ for $L=3.0$ and $L=4.0$ in $H=6.8$ T (red curve) and zero (black curve) applied field along the $c$ direction, taken at $T=2$ K. This data was taken at RITA-II, PSI. For each scan the data of five blades have been combined leading to an uncertainty in $L$ of $0.1$ r.l.u.. Errorbars are smaller than the markersize. 
The field data are shifted upwards by a constant offset for clarity. 
(c) The fitted peak position $H_{\rm IC}$ for $6.8$ T field data as a function of $L$. 
The color inset shows the intensity of the structural scattering signal around $(2~0  ~0)$.}
\label{fig:RITAdata}
\end{figure*}

The PANDA experiment was performed in zero field after a quench-cooling of the crystal by 
 $4$~K/min. The experimental setup was $E_i = E_f =5.0$~meV and we used 60$^\prime$ collimation before and after the sample. A Be filter was placed between monochromator and the sample.

In the experiments at FLEXX and RITA-II horizontal field magnets were used.
The experimental setup on RITA-II was $E_i = E_f =4.6~$meV and the 9-blade analyzer arranged in the monochromatic imaging mode.~\cite{RITA} We used a 80$^\prime$ collimation before the sample and a Be filter with radial collimation after the sample. The sample was mounted in a $6.8$~T horizontal field cryomagnet and data was taken in $6.8$ T field applied along the $c$-axis as well as in zero field. 
To improve data statistics, we performed a similar experiment at FLEXX with the sample placed in a 6~T horizontal magnet with the same orientation as in the RITA-II experiment. The FLEXX experiment was performed with energies $E_i = E_f = 5.0~$meV and we used 60$^\prime$ collimation
between monochromator and sample as well as between sample and analyzer. Second order contamination from the monochromator was eliminated by a velocity selector.
In both the RITA-II and FLEXX experiments the same slow sample cooling of 1K/min was performed.
We studied the magnetic order by scanning through the magnetic ordering vector ${\bf Q}_{\rm IC}$ at $T=2$ K by rotating both the sample and magnet, keeping the magnetic field along the $c$-axis. In some of the scans the background contribution was estimated by performing similar scans at $T=40$~K, where the magnetic order is absent.

\section{Results}
\label{section_res}

\begin{figure}[b!]
\includegraphics[clip=true,width=0.9\columnwidth]{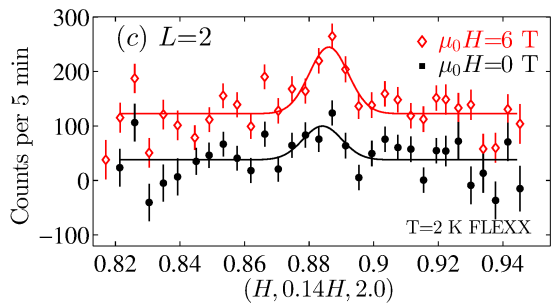}
\caption{(color online) Measured peak intensity at $T=2$ K for the IC position $(H~0.14H~2.0)$ in $6$ T field (red curve) and zero field (black curve). Point-wise background subtraction  has been performed, using 40~K data as background. This data was taken at FLEXX, HZB. The field data is shifted upwards by a constant offset for clarity.}
\label{fig:FLEXXdata}
\end{figure}

Figure~\ref{fig:RITAdata} (a,b) show the raw data taken on RITA-II through the IC position ${\bf Q}_{\rm IC}$ at different magnetic fields for $L=3$ and $L=4$, respectively. It is seen that for the zero field data the peak intensity is roughly the same at $L=3$ and $L=4$, whereas the measurements in an applied field show a higher intensity at $L=4$ than at $L=3$. We fit the raw data to a single Gaussian on a sloping background keeping the width of the peak constant. The peak width is resolution limited and corresponds to a large in-plane correlation of $\xi_{\rm in-plane} \geq 120$~\AA, consistent with earlier measurements finding resolution limited correlations in the $(a,b)$ plane.~\cite{Romer13} Due to the use of imaging mode the RITA-II experiment amounts to almost 80 individual scans. In Figure~\ref{fig:RITAdata} (a,b), the data of five blades have been combined for an integration range of $\Delta L=0.1$ r.l.u.. For the individual scans the fitted center position of the peaks varies 
within $\approx 0.002$ r.l.u. around the mean value of $H_{\rm IC} = 0.887$. Further inspection of the fitted peak center shows a clear modulation as a function of $L$ as shown in Fig.~\ref{fig:RITAdata} (c) with $H_{\rm IC}=0.888$ for even $L$ and $H_{\rm IC}=0.886$ for odd $L$. Although significant, the variation in the fitted peak center is smaller than the resolution limited width of $0.006$ r.l.u. and we cannot resolve the signal into two separate peaks. We will later see that the clear modulation of the peak center provides evidence that the IC signal is twinned. In the inset of Fig.~\ref{fig:RITAdata}~(c) the twinning of the structural peak at $(2 ~0~0)$ is depicted, showing that three distinct structural domains are visible. 

In Fig.~\ref{fig:FLEXXdata} we show scans at $L=2$ from the independent experiment on FLEXX. In this experiment we did a detailed map of the background measured by sample rotations above the magnetic ordering temperature at $T=40$ K. For these data, a point-wise background subtraction has been done. 
The figure clearly shows the effect of a 6 T magnetic field; the IC
magnetic signal is roughly doubled. This is in agreement with the enhancement of the $L=0$ signal in $7$~T, observed in Ref.~\onlinecite{Romer13}.
\begin{figure}[t!]
\includegraphics[clip=true,width=0.95\columnwidth]{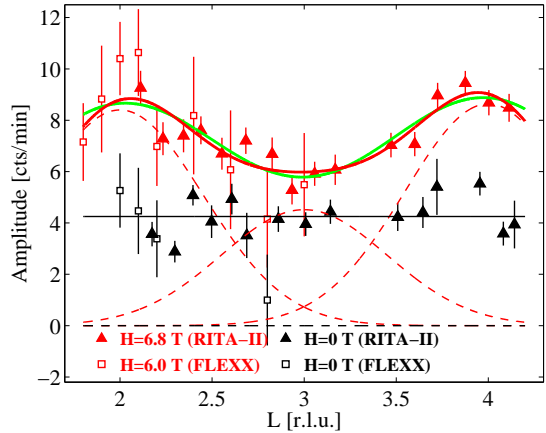}
\caption{(color online) Background-subtracted elastic response at the IC position $(0.887(2),~0.124(1),~L)$ versus $L$ in zero field, a $6.0$ T field (FLEXX data), and a $6.8$ T field (RITA-II data). 
The $y$-axis label corresponds to the counting rate at the RITA-II experiment. The FLEXX data has been scaled by a constant factor to compare with the RITA-II data. The red solid line corresponds to a fit to three Gaussians with the same width and fixed centers at $L=2,3$ and $4$, the three Gaussians are shown separately by the red dashed lines. The green line is a fit to two Gaussians with the same width and fixed centers at $L=2$ and $L=4$.}
\label{fig:Ampl_vs_L_RITAandFLEXX}
\end{figure}

Now we turn to the main purpose of the study, which is to map out the full $L$-dependence of the IC signal in field and zero field. We did several scans similar to those shown in Figs.~\ref{fig:RITAdata} (a,b) and~\ref{fig:FLEXXdata} for $L$ in the range $1.8$ to $4.15$.
The results are summarised in Fig.~\ref{fig:Ampl_vs_L_RITAandFLEXX} for both the RITA-II and FLEXX experiments. In zero field the measured IC signal is flat as a function of $L$, i.e. there is no observable interplanar correlations when the sample has been cooled down slowly.
The effect of applying a magnetic field perpendicular to the copper oxide planes is twofold. First, weak correlations between the CuO$_2$ planes develop in agreement with the observations of Lake {\it et al.} in LSCO $x=0.11$.~\cite{Lake05} Secondly, and more pronounced, an overall enhancement of the magnetic signal takes place for all values of $L$. A fit of the field data to three Gaussians with same width and fixed centers at $2,3$ and $4$ gives a broad modulation with $\sigma=0.45(4)$ r.l.u. This corresponds to a very short correlation length of 4 \AA, smaller than the distance between neighboring CuO$_2$ planes. 

As a measure of the true enhanced intensity we integrate the signal measured in field along $L$ and compare to the $L$-integrated zero-field signal. From Fig.~\ref{fig:Ampl_vs_L_RITAandFLEXX} we get an $L$-integrated enhanced intensity of $77.1(8)$\%. 
For a comparison we calculate the enhancement effect at $L=2$ from Fig.~\ref{fig:Ampl_vs_L_RITAandFLEXX} and get $99.4(6)$\%. The latter corresponds to the effect which would be estimated from a measurement with the current crystal aligned in the $(a,b)$ plane. 

\begin{figure*}[t!]
\centering
\includegraphics[clip=true,width=0.35\linewidth]{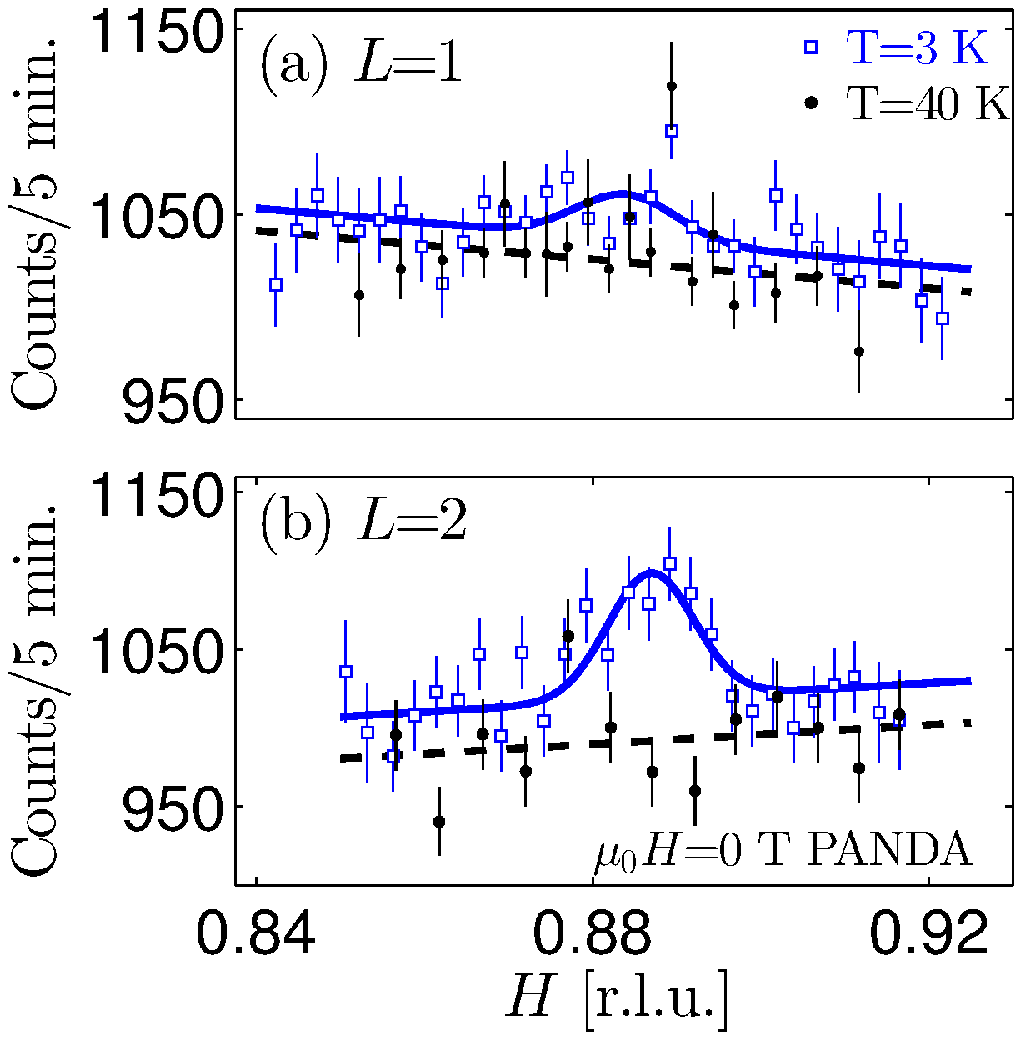}
\includegraphics[clip=true,width=0.45\linewidth]{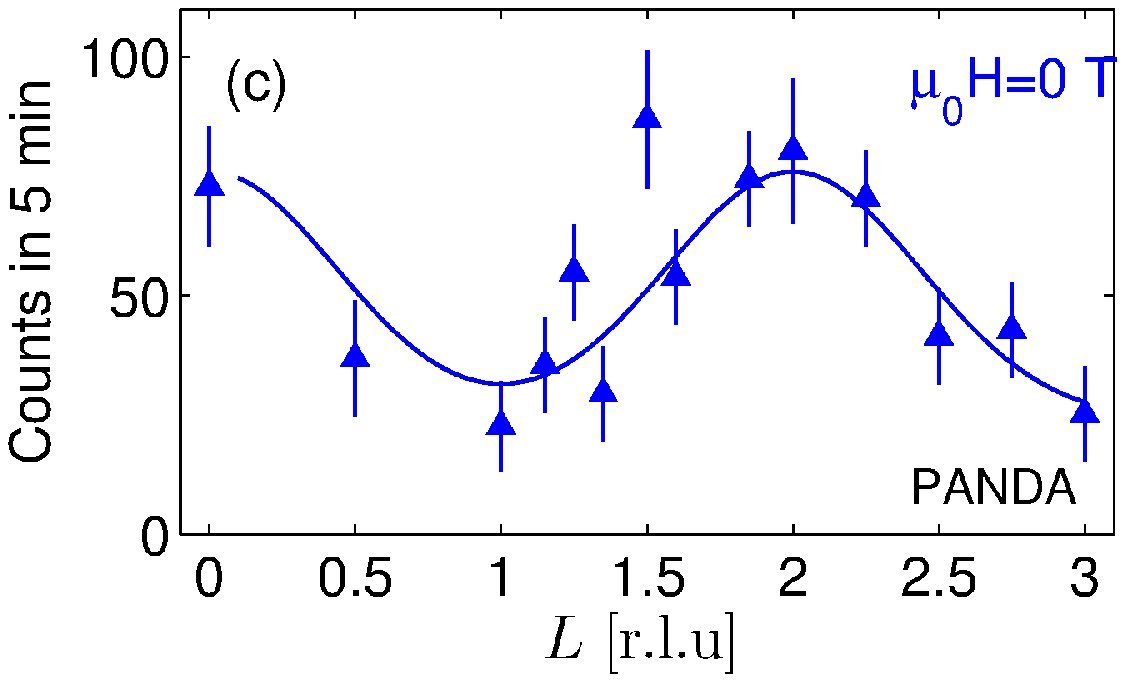}
\caption{(color online) (a-b) Scans through $(H~0.14H~L)$ at $T=3$ K (blue open squares) and $T=40$ K (black points) in zero field for $L=1$ (a) and $L=2$ (b). (c) $L$-dependence of the incommensurate magnetic signal above background in zero field measured at PANDA.}
\label{fig:Ampl_vs_L_PANDA}
\end{figure*}

Finally, in Fig.~\ref{fig:Ampl_vs_L_PANDA} we show the results of the PANDA experiment, which was done at zero field, but under different experimental conditions since the crystal was quenched-cooled by 4~K/min. We observe a small correlation between the CuO$_2$ planes even when no magnetic field is applied. In this case a fit to a Gaussian function at even $L$ gives a width of $\sigma_L=0.4(1)$ r.l.u. similar to the broad modulation observed in field. This leaves us with the interesting observation that quench-cooling has the same qualitative effect of enhancing inter-planar correlations as the application of an external magnetic field in the $c$-direction.

\section{Discussion}
\label{section_dis}
In the previous section we reported that $c$-axis correlations are absent when the system is cooled slowly from room temperature to base temperature. Further, we observed that a magnetic field as well as quench-cooling can lead to the development of clear, but short-range, $c$-axis correlations.
Now we turn to a discussion of the magnetic signal and how it is affected by an applied magnetic field and quench-cooling the system. Finally, we discuss our experimental findings in the framework of theoretical microscopic models.

\subsection{Peak position of the magnetic signal}
\label{sec:discus_signal}
We observe an IC signal at all values of $L$ in field as well as in zero field. In field the position is at $H=0.887(2)$ with a weak $L$ dependence as shown in Fig.~\ref{fig:RITAdata}~(c). Under the assumption that this IC signal belongs to the dominating structural twin, the mean IC position corresponds to a displacement angle of $\theta=2.7^\circ$ away from the high symmetry axis of the underlying CuO$_2$ plane. This is consistent with the findings of Kimura {\it et al.}\cite{Kimura00} on a crystal of similar doping. In zero field the mean peak position is the same as in field with a slightly larger uncertainty due to the very weak signal. In zero field there is no discernible variation in peak position along $L$. 

The signal structure along the $L$ direction for different twin components of the IC order in LSCO has not been measured previously. Our expectations stem from the parent compound La$_2$CuO$_4$\cite{Vaknin87,Endoh88} and super-oxygenated La$_2$CuO$_{4+y}$.\cite{Lee99} 
The parent compound La$_2$CuO$_4$ displays long-range three-dimensional antiferromagnetic order with spins aligned mainly along the orthorhombic $b$-axis.~\cite{Vaknin87,Endoh88} With this spin structure the magnetic signal is peaked at {\em even} $L$ for scattering at $(1~0~L)$ and at {\em odd} $L$ for scattering at $(0~1~L)$.
In the work by Lee {\it et al.}~\cite{Lee99} the three-dimensionality of the IC signal in super-oxygenated La$_2$CuO$_{4+y}$ was investigated. A clear splitting of the incommensurate signal allowed measurements of IC signals centered at $(1~0~L)$ and $(0~1~L)$ simultaneously.  Lee and coworkers report a broad $L$-dependence of the IC signals; for the IC signal centered at $(1~0~L)$ the peak is at even $L$ while it occurs for odd $L$ for the IC signal centered at $(0~1~L)$. 
Thus in La$_2$CuO$_{4+y}$ the arrangement of the spins in the inter-planar direction bear resemblance to the spin arrangement in the parent compound with the spins being correlated across two to three CuO$_2$ planes.~\cite{Lee99}

In our case the orthorhombicity of the crystal is much smaller than in La$_2$CuO$_{4+y}$ which makes it harder to identify a possible twinning of the IC signal.
From scans through the structural peaks three structural domains are visible as shown in the inset of Fig.~\ref{fig:RITAdata}~(c). The sub-dominant second order peak close to $(1~0~0)$ is displaced from the dominant peak by $0.8$ \% in both directions. This corresponds to a rotation of $0.22^\circ$ between the two twins. In Ref.~\onlinecite{Kimura00} the twin splitting was reported to be  $\sim 0.3^\circ$.

By a rotation of $~0.22^\circ$ and a rescaling of the vector length, the structural peak of the dominant twin is brought to the same position as the structural peak of the sub-dominant twin. Performing the same transformation on the IC peak at $(0.888,~0.124,~0)$  we might expect an IC twin at the position $(0.879,~0.126,~0)$. This predicts a larger deviation in peak center than expected from the lower value of the fitted peak center shown in Fig.~\ref{fig:RITAdata}~(c) for $L=3$ which is $H=0.886$. Note, however, that the difference in peak position is of the order of the resolution limitation of $0.006$ r.l.u.. Therefore, it is likely that resolution limitations cause both IC twin signals to contribute for all values of $L$ with one twin $(0.888,~0.124,~0)$ dominating at even $L$ and the other twin dominates for odd $L$.

We measure the IC signal centered at $(1~0~L)$ defined with respect to the dominant structural domain. If the magnetic structure of LCO is partially maintained also in its Sr-doped offspring we expect inter-planar correlations from the dominant twin to be peaked at even $L$. By resolution effects the IC signals from the two sub-dominant structural twins are also picked up, and since these are both centered at $(0~1~L)$ part of the signal will be peaked at odd $L$.

As seen in Figs.~\ref{fig:Ampl_vs_L_RITAandFLEXX} and~\ref{fig:Ampl_vs_L_PANDA}, the largest component of the signal shows a peak at even $L$, but as proposed in Fig.~\ref{fig:a3scan}~(b) we might expect part of the signal to be peaked at odd $L$. If the signal was peaked only at even $L$ a fit to two Gaussians (green curve in Fig.~\ref{fig:Ampl_vs_L_RITAandFLEXX}) is appropriate. However, if part of the signal is peaked at odd $L$, the data should rather be fitted to three Gaussians as shown in Fig.~\ref{fig:Ampl_vs_L_RITAandFLEXX} by the red curve. In this interpretation the correlations along the $L$ direction is equally well-developed for all twins, so the width of the Gaussian at $L=3$ is equal to the width at $L=2$ and $4$. Since the twin which shows correlations peaked at even $L$ is more pronounced than the twins with correlations peaked at odd $L$ we see only the peaks at even $L=2$ and $4$ whereas the peak at $L=3$ is masked by the tails of the other two. Due to the broadness of the signals the peak 
position is an average over both twins at all values of $L$ which is probably why we observe only a very small shift in the peak center at odd $L$ compared to even $L$. Our observations are consistent with this picture.

\subsection{Field induced $c$-axis correlations}
From the two independent field experiments done at RITA-II and FLEXX we find that the IC signal at $(0.887~0.124~L)$ is enhanced by a magnetic field at all values of $L$, but in particular for even $L$. The results of both experiments agree on the magnitude of the magnetic field enhancement at even $L$ which is close to a factor of two in both cases. 

Although the FLEXX experiment provided less data points, it is clear from Fig.~\ref{fig:Ampl_vs_L_RITAandFLEXX} that both experiments agree on the development of weak inter-planar correlations in a field as reflected in the $L$-dependence of the amplitude of the IC intensities. The peaks which are centered at even values of $L$ are very broad and the correlation length is smaller than the inter-planar distance of $6.5$ \AA. This deviates from the correlation length of more than $10$ \AA ~as found in Ref.~\onlinecite{Lake05} for LSCO $x=0.11$. We note that the correlation length determined in Ref.~\onlinecite{Lake05} is likely uncertain due to the sparse data. 
However, we cannot rule out a real difference in the field-induced inter-planar correlation length between these two crystals of different doping levels. Since our crystal displays enhanced magnetic order in zero field compared to smaller doping values, it might be harder to change the inter-planar correlations by a field.

The primary effect of an applied magnetic field is to enhance the magnetic signal either by enlarging the magnetic volume fraction or the ordered magnetic moments. Our neutron scattering experiment does not allow for distinguishing between these two possible scenarios. 
Muon spin rotation studies on the same crystal show that magnetic order is present throughout the entire volume of the sample with a resolution of 20 \AA~given by the range of sight of the muons inside the sample~\cite{Chang08,JacobThesis} and we conclude that the main effect arises from enlargening of the ordered magnetic moments.

In addition, development of weak inter-planar correlations occurs as a response to the applied magnetic field. 
As a consequence the field effect reported in the literature on magnetism in the cuprates that has been measured in the usual configuration $L=0$ must be re-interpreted. A common interpretation of the field effect has been that it increases either the magnetic moments or the magnetic volume fraction. In the $L=0$ setup this effect is either over-estimated or under-estimated, depending on whether the dominant IC peak belongs to a ``$(1 ~0 ~0)$'' or ``$(0~1~0)$'' IC quartet. For our crystal measuring at the IC peak at $(0.887, ~0.124,~0)$ with $L=0$ would cause an over-estimation of the field effect. At even $L$ the amplitude enhancement is roughly $100$\% whereas the real increase of  magnetic order measured from the $L$-averaged intensity is only $77$\%. Thus the field effect is over-estimated by 30\%. 

\subsection{Cooling-induced $c$-axis correlations}
Another new finding in this work is the fact that inter-planar correlations are also found in zero field under different experimental conditions where the crystal is quench-cooled from room temperature down to base temperature below 4~K. In contrast to the experiments on RITA-II and FLEXX, where the structural phase transition as well as the ordering of possible excess oxygen were traversed slowly, both processes were passed extremely fast in the PANDA zero-field experiment. This is likely to have resulted in finite inter-planar correlations as evident from Fig.~\ref{fig:Ampl_vs_L_PANDA}, qualitatively similar to the result of an applied magnetic field.

To understand this behaviour, we first compare our sample to super-oxygenated crystals where the excess oxygen order in  a three-dimensional structure upon slow cooling. 
Lee {\em et al.}\cite{Lee04} showed that fast cooling leads to an oxygen-disordered state displaying enhanced spin-density-wave (SDW) order compared to the oxygen-ordered state. In fact, in this work it was observed that disordering the excess oxygen has the same enhancement effect of the SDW signal as the application of a 7.5 T field. However, it remains unknown how disordered oxygen ions or applied magnetic field affect the magnetic correlations between the CuO$_2$ planes.

In general, the cooling history is known to be important for the strength of the IC signal and reproducibility of the signal strength is obtained only when cooling slowly through the transition temperature of the oxygen-ordered state. Sensitivity to the cooling procedure is also likely to exist in the Sr-doped samples, even though this has been less discussed in the literature. Following this line of thought we put forward two possible explanations for our observations in 12 \% doped LSCO.

First, our sample might have a small amount of excess oxygen since this is not easily avoided during crystal growth. Fast cooling through the temperature regime where ordering of possible excess oxygen takes place, which occurs down to $\sim 180$~K, might cause random positions of the excess oxygen ions. Such impurities could act as pinning centers enhancing the magnetic correlations between the planes.

Second, the crystal has a tendency towards domain formation when cooled through the structural transition  from high-temperature-tetragonal (HTT) to low-temperature-orthorombic (LTO) at $T_{\small \rm HTT-LTO}=255$~K.~\cite{Chang08} Domains form in the ($a,b$) plane due to the small difference of these two lattice constants in the orthorhombic phase. 
When slowly cooled, the crystal shows at three discernible domains out of four possible. The fast cooling procedure done in the PANDA experiment might change this domain formation and result in a different twin pattern, which is likely to contain a larger number of smaller domains than with slow cooling. 
These domains could act as pinning centers along the inter-planar direction since they penetrate the sample through this direction. Such a pinning center that penetrates the sample perpendicular to the CuO$_2$ planes might result in enhanced correlations along this direction in a similar way as pinning centers arising from vortices in an applied field.

\subsection{Theoretical scenario}

Theoretically, the slowing down and subsequent pinning of static magnetic order by disorder sites and twin boundaries,\cite{Tsuchiura,Wang,Chen,Kontani,Harter,Andersen07,Andersen10,Christensen,Tricoli} and vortices\cite{Arovas,Andersen00,Demler01,Chen02,Zhu,Takigawa,Andersen09,Andersen11} has been previously discussed extensively in the literature. From the microscopic studies, it is clear that the modulations of charge density and/or electron hopping amplitudes induced by impurities and twin boundaries can lead to local magnetic instabilities which nucleate magnetic order in the vicinity of the perturbing sites. The vortices, on the other hand, typically induce local magnetic order due to the suppressed superconducting gap and an associated enhanced local density of states near the Fermi level in the cores. It has been shown that even in strongly disordered situations vortices enhance the in-plane magnetic moments.\cite{Schmid10,Andersen11} To the best of our knowledge, the out-of-plane induced magnetic order by disorder or by 
vortices has not been described by microscopic models, and constitute an interesting future study. For the pure superconducting system, flux lines along the $c$-axis should lead to substantially enhanced spin correlations along $L$.\cite{Lake05} The weak coupling between the CuO$_2$ planes will lead to short-ranged vortex-induced magnetic order, but the extremely short $c$-axis correlations found here points to additional effects. Certainly, the full magnetic volume fraction already in zero field indicates that there is hardly any "room" for vortices to induce coherent spin correlations along the $c$-axis. Instead, the vortices lead to local enhancements of the magnetic moments and presumably adapt to the many pre-existing pinning centers and strongly meander along $c$, leading to only very weak $c$-axis field-induced correlations in qualitative agreement with our observations. Within this scenario, higher doped samples with less static magnetic order in zero field, should lead to longer-ranged and more pronounced $c$-axis correlations in the presence of a magnetic field.

\section{Conclusions}
We have studied the field dependence of the inter-planar magnetic correlations in La$_{1.88}$Sr$_{0.12}$CuO$_4$. 
The primary effect of an applied magnetic field is an enhancement of the magnetic moments. Further, there is an effect of increased inter-planar correlations in the presence of an applied field. The inter-planar correlation length is very small and imply correlations only between neighbouring planes. This indicates that the magnetic order is already strongly pinned by impurities in the sample and that vortices tend to bend rather than go perpendicular to the CuO$_2$ planes on the way through the sample.
We observe that a fast cooling procedure results in the same feature as application of a magnetic field, namely development of weak inter-planar correlations. Two possible scenarios - caused by excess oxygen or micro-domains - could lead to pinning of the magnetic order between the CuO$_2$ planes thereby explaining why a quench-cooled system behaves similarly as a system subjected to an external field. 

\section*{Acknowledgements}
We are greatful for the access to neutron beam time at the neutron facilities at NCNR-NIST, BER-2 at Helmholtz-Zentrum Berlin, FRM-2 at
MLZ Garching, and  SINQ at the Paul Scherrer Institute.  This project was supported by the Danish Council for Independent Research through DANSCATT. B.M.A. acknowledges support from the Lundbeckfond (fellowship grant A9318).


\begin{thebibliography}{0}
\bibitem{RealdopingofBellaCryst} The doping of the crystal in Ref.~\onlinecite{Lake05} was later shown to be closer to $x=0.11$ rather than $x=0.10$ from a measurement of the transition temperature from HTT to LTO, R. Gilardi Ph.D. Dissertation~Ref.\onlinecite{GilardiThesis}.
\end{thebibliography}
\end{document}